\documentclass[journal=jacsat,manuscript=article]{achemso}
\pdfoutput=1
\usepackage{graphicx}

\author{Kurt Stokbro}
\email{kurt.stokbro@quantumwise.com}
\affiliation{QuantumWise A/S,
 N{\o}rre S{\o}gade 27A, 1.\ th,
  DK-1370 Copenhagen K, Denmark}

\title{First-principles modelling of molecular single-electron transistors}

\begin{document}
\begin{abstract}
 We present a first-principles method for calculating the charging
 energy of a molecular single-electron transistor operating in the
 Coulomb blockade regime. The properties of
 the molecule are modelled using density-functional theory,
  the environment is described by a continuum model, and the interaction
  between the molecule and the environment are included through the Poisson
  equation. The model is used to calculate the charge stability
  diagrams of a benzene and C$_{60}$
  molecular single-electron transistor.
\end{abstract}

\section{Introduction}

The use of non-equilibrium Greens functions (NEGF) in
connection with density-functional theory (DFT)\cite{Lang1995,Xue2002,Brandbyge-2002-PRB,Taylor2001} or semi-empirical
models\cite{MaJo97,CoCeSa99,Cerda2000,EmKi01,Zahid2005,KienleI2006,KienleII2006,
stokbro2010} has been highly successful in modelling coherent transport in various types of molecular
junctions. However, in the case of molecular single-electron
transistors (SET), the transport is incoherent\cite{bjornholm}, and another approach is
needed. Kaasbjerg \textit{et al.} introduced a semi-empirical model
for simulating the properties of molecular SETs. In particular they showed the importance of including
renormalization of the molecular charge states due to the
polarization of the environment.

In this paper we extend this framework to be
included within a density-functional theory description of
molecular SET's operating in the coulumb blockade regime. We use the model to calculate the charging energy of
benzene and C$_{60}$ in an electrostatic environment resembling a
molecular SET geometry. We calculate
the charging energy as function of an external gate potential, and
from this we obtain the charge stability diagram of the
two respective molecules.

The outline of the paper is as follows. In the first section we
describe the basics of an SET, and in the
following section the DFT framework for modelling the device. We next
present calculations of the properties of benzene
and C$_{60}$ in an electrostatic environment, and in the final
section we summarize the results.

\section{Basic theory of a molecular single electron transistor}

\begin{figure}[tbp]
\begin{center}
\includegraphics[width=9cm]{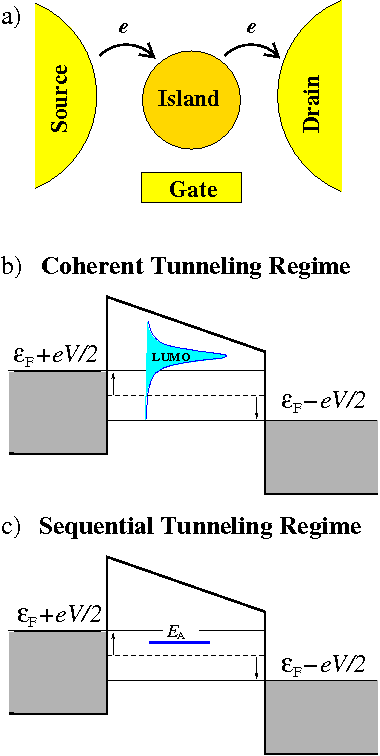}
\end{center}
\caption{(Color online) a) Schematics of a nanoscale transistor. Electrons propagate from the source to the
        drain through an island. The energies of the electronic states on the island can be
        controlled by an electrostatic gate. b) In the strong coupling regime, the electron propagates coherently through the device, and
        the electronic states of the island have a finite lifetime and
        therefore a finite width. c) In
        the weak coupling regime, the electron localizes on the island and propagates by
        sequential tunneling. The electronic states on the island are in this case
        discrete.}\label{fig:set}
\end{figure}

\ref{fig:set}a schematically illustrates the geometry of a nanoscale molecular
transistor. The geometry consists of metallic source and drain
electrodes, and a molecular island coupled with the two electrodes.
Electrons can propagate from source to drain through the island.

If the island is strongly coupled with the source and drain electrodes,
the electrons will stay a very short time on the island, and cannot localize but will
move coherently through the system. This is the regime where we
can use the coherent transport model for simulating the
electrical properties of the system. This is the situation shown
in \ref{fig:set}b, and we note that a
current can flow through the system even when the island does not have any
electronic states within the
bias window, as illustrated
in the figure by the finite lifetime-broadened lowest unoccupied
molecular orbital (LUMO), which the carrier can use for
propagating from left to right.

In this paper, we will investigate the regime where the
island is weakly coupled
with the electrodes. In this case, the electron tunnels from the source to the
island and stays there for sufficiently long time to localize. The electron thereby
loses all information about its original quantum state. Thus, the subsequent tunneling
process from the island to the drain electrode will be independent
from the tunneling process into the island. This transport mechanism is referred to
as sequential tunneling and is depicted in
\ref{fig:set}c. Electron transport is in this case only possible if the island has an electronic
level within the bias window, as illustrated by the
electron affinity level
(EA) in \ref{fig:set}.
As in the coherent case, the position
of the molecular levels and in particular the electron affinity level can be adjusted by an external gate potential, and by
appropriate tuning, the island can in this regime thus be opened or closed for transport. For
source-drain voltages below the charging energy of the island, there will only be
one energy level within the bias window, and the system will work as a single
electron transistor, as desired for our present discussion.

\subsection{The energy balance in the weak coupling regime}

In the following we will focus on the weak coupling regime where the transport is described by
sequential tunneling. We introduce the function $ E^{\rm island}(N)$,
which gives the total energy of the island as function of the number of electrons on the
island. We also introduce similar energy functions for the source and drain electrodes,
$E^{\rm source}(N)$ and $E^{\rm drain}(N)$.

For the electron to move from the source electrode onto the island, the electron
must have a lower energy on the island, i.e.\
\begin{equation}
  E^{\rm source}(M) + E^{\rm island}(N) \ge  E^{\rm source}(M-1) + E^{\rm island}(N+1),
\end{equation}
where $N$ and $M$ are the initial number of electrons on the island and in the source electrode, respectively.

Moreover, in order to move the electron from the island to the drain electrode, it must have a
lower energy in the drain electrode,
\begin{equation}
  E^{\rm drain}(K) + E^{\rm island}(N+1) \ge  E^{\rm drain}(K+1) + E^{\rm island}(N),
\end{equation}
where $K$ is the initial number of electrons in the drain electrode.

The maximum energy of the electron in the source electrode is $ -W + e V/2$,
where $W$ is the work function of the electrode and $V$
the applied bias. Assuming that the electron with maximum energy tunnels onto
the island, then we have
\begin{equation}
  E^{\rm source}(M) - E^{\rm source}(M-1) =  -W + e V /2.
\end{equation}
Using the above tunneling criterion, we obtain the condition
\begin{equation}
  -W + e V /2 +      E^{\rm island}(N) \ge     E^{\rm island}(N+1).
\end{equation}

Similarly, $ -W - e V/2 $ is the minimum energy of an electron in the drain
electrode, and thus
\begin{equation}
  E^{\rm island}(N+1) \ge   -W - e V /2 +      E^{\rm island}(N).
\end{equation}

The requirement for a current to flow in the device is therefore
\begin{equation}\label{eq:charge-stability}
  e |V| /2     \ge \Delta E^{\rm island}(N)+W \ge  -e |V| /2,
\end{equation}
where $ \Delta E^{\rm island}(N) = E^{\rm island}(N+1)-E^{\rm island}(N)$
is the charging energy of the island.

In the following we will calculate the charging energy of two
different molecular SETs and use
\ref{eq:charge-stability} to obtain the so-called charge
stability diagram, which shows the number of charge states inside the
bias window as function of the gate and source-drain voltages.

\subsection{Total energy of a molecule in an electrostatic environment}

In this section we will discuss the calculation of the total energy of
a molecule in an electrostatic environment. All calculations are performed
using the commercial software package Atomistix ToolKit (ATK)\cite{ATK2010.02}.
The DFT model in ATK is based on pseudopotentials with numerical localized
basis functions using the method outlined by Soler {\it et al.}\cite{SIESTA2002}.
In this framework, a compensation charge  $\rho^\mathrm{comp}_i({\bf r})$ is
introduced for each atomic site. The compensation
charge has the same charge $Z_i$ as the pseudopotential, and is used to
screen the electrostatic interactions.

We now introduce the electron difference density
\begin{equation}
\delta n({\bf r}) = n({\bf r}) - \sum_i \rho^\mathrm{comp}_i({\bf r}),
\end{equation}
where $n({\bf r})$ is the total charge density of the system. We also
introduce the screened (``neutral atom'') local pseudopotential
\begin{equation}
V^\mathrm{NA}_i({\bf r}) = V^\mathrm{loc}_i({\bf r})-
\int \frac{\rho^\mathrm{comp}_i({\bf r'})}{|{\bf r}-{\bf r'}|} d{\bf r'},
\end{equation}
where $V^\mathrm{loc}_i$ is the local pseudopotential at site $i$.

Following Ref.~\citenum{SIESTA2002}, we rearrange the terms to obtain the
DFT total-energy functional
\begin{equation}
E[n] = T[n] + E^\mathrm{xc}[n] +\frac{1}{2} \int \delta V^\mathrm{H}({\bf r}) \delta n({\bf r}) d
{\bf r} +  \int \sum_i V^\mathrm{NA}_i({\bf r}) \delta n({\bf r})
d{\bf r} + \frac{1}{2} \sum_{ij} U_{ij},
\label{eq:tot}
\end{equation}
where $T$ is the one-electron kinetic energy, $E^\mathrm{xc}$ the exchange-correlation
energy, and the three last terms are the rearranged electrostatic
terms. The first of these is the relative Hartree energy, and is
obtained from the difference Hartree potential $\delta V^\mathrm{H}({\bf r})$
which is calculated by solving the Poisson equation for the
difference density $\delta n({\bf r})$. The next term describes the relative external
energy, while the last term contains all the electrostatic interactions which do not depend on
the electron density. The last term is calculated from
$V^\mathrm{loc}_i$, $\rho^\mathrm{comp}_i$, and $Z_i$, and since the Poisson equation
is linear, this term can be decomposed into the pair-potential $U_{ij}$.

\begin{figure}[tbp]
\begin{center}
\includegraphics[width=\linewidth]{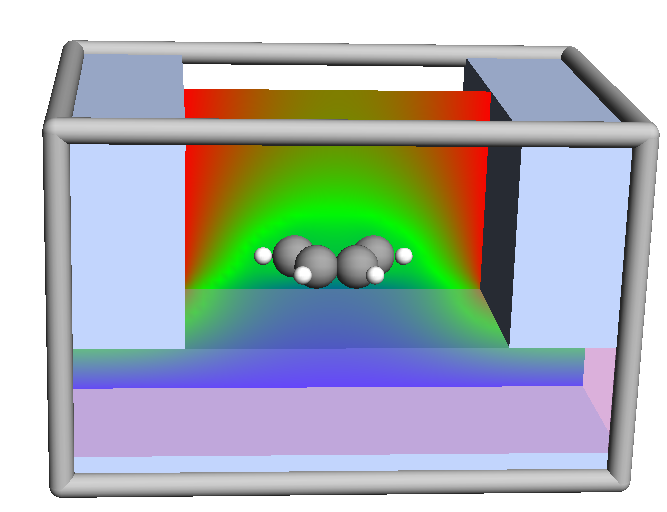}
\end{center}
\caption{(Color online) A benzene molecule in the SET environment 
considered in this paper. The electrostatic environment models 
two metallic electrodes (source and drain) on top of a dielectric 
substrate with a metallic back-gate. The contour plot shows the 
induced electrostatic potential for a gate voltage of 2~V and 
zero source-drain bias.}\label{fig:benzene_set}
\end{figure}

We next extend the total-energy functional to include interactions with
a number of dielectric and metallic regions surrounding the molecular
system. \ref{fig:benzene_set} illustrates a typical molecular single electron
transistor geometry where a benzene
molecule is positioned on top of a dielectric material and surrounded
by three metallic electrodes. Within the metallic regions the
potential is fixed to the applied voltage on each respective electrode.

Solving the Poisson equation without the molecule present,
we obtain the external potential from the electrostatic environment,
\begin{equation}
  \label{eq:poisson}
  - \nabla \cdot \left[ \epsilon(\mathbf{r})\, \nabla  V^\mathrm{ext}(\mathbf{r}) \right] = 0.
\end{equation}
Adding the molecule to the geometry, we again solve the Poisson
equation, but now with the electron difference density on the molecule, to obtain the
total difference Hartree potential
\begin{equation}
  \label{eq:difference_poisson}
  - \nabla \cdot \left[ \epsilon(\mathbf{r})\, \nabla  \delta
    V^\mathrm{H+ext}(\mathbf{r}) \right] = \delta n(\mathbf{r}).
\end{equation}

Finally, we define the molecular part of the total difference Hartree
potential
\begin{equation}
 \delta   V^\mathrm{H}(\mathbf{r}) =   \delta  V^\mathrm{H+ext}(\mathbf{r}) - V^\mathrm{ext}(\mathbf{r}),
\end{equation}
and this is the difference Hartree potential which enters into \ref{eq:tot}.
Following Neugebauer and Scheffler\cite{NeSc92}, we add the energy
contribution from the external field through the term
\begin{equation}
\Delta E = \int V^\mathrm{ext}({\bf r}) n({\bf r}) d ({\bf r})-\sum_i
V^\mathrm{ext}({\bf R}_i) Z_i,
\label{eq:ext}
\end{equation}
where ${\bf R}_i$ is the position of site $i$, and $Z_i$ the valency
of the pseudopotential at site $i$.

Thus, the total energy is given by adding the contributions from \ref{eq:tot} and \ref{eq:ext}.
Note that this is only exact if the last electro-static pair-potential term
in \ref{eq:tot} is
unaffected by the electrostatic surroundings, which is the
case if the compensation charge and the screened local pseudopotential
do not overlap with the metallic and dielectric regions.

\section{Results}

We will now present results for the charging energy of a benzene and
C$_{60}$ molecule in an SET geometry. To obtain some
reference energies, we will first calculate the charging
energies of the isolated molecules. We obtain the charging energy
by performing self-consistent calculations for the $N$ and $N+1$ charge
states of the isolated molecule and subtracting their total energies.
For the calculation we use non-polarized DFT in the local density approximation
(LDA)\cite{Perdew1981} and expand
the wavefunctions in a double-zeta polarized (DZP) basis
set. The molecular geometries where obtained by relaxing the molecules in the
neutral state.

\begin{table}
  \caption{Experimental ionization (I) and affinity (A) energies
    for benzene and C$_{60}$ molecules compared with theoretical
    DFT-LDA values. The theoretical values denoted ``isolated'' are
    obtained from the total energies of the charged isolated
    molecule. The ``SET'' values are obtained by calculating the total
    energy of the charged molecule in the
    electrostatic surrounding illustrated in \ref{fig:benzene_set},
    at zero gate and source-drain bias.}
  \label{tab:total_energy}
  \begin{tabular}{lllll}
    \hline
    benzene & $E_\mathrm{I}^{+1}$ & $E_\mathrm{I}$ & $E_\mathrm{A}$ & $E_\mathrm{A}^{-1}$ \\
    \hline
    Exp. & &9.25\cite{benzene_ion} & & \\
    isolated & 15.73 & 9.15 & $-2.34$ & $-8.39$ \\
    SET  & 7.70 & 5.41 & $-2.26$ & $-4.88$ \\
    \hline
    C$_{60}$ & $E_\mathrm{I}^{+1}$ & $E_\mathrm{I}$ & $E_\mathrm{A}$ & $E_\mathrm{A}^{-1}$ \\
    \hline
    Exp. & 11.33\cite{c60_ion}& 7.65\cite{c60_ion} & 2.65\cite{c60_a}& \\
    isolated & 10.09 & 6.84  & 1.90  & $-1.32$ \\
    SET  & 7.24 & 5.89 & 2.85 & 1.53 \\
    \hline
  \end{tabular}
\end{table}

\ref{tab:total_energy} shows the calculated
charging energies. For benzene there is excellent agreement with the experimental
results, while the results for C$_{60}$ are about 1~eV too small. To investigate
possible origins of this discrepancy we performed
calculations where C$_{60}$ was allowed to spin polarize and
relaxed the molecule also in the charged state. We found that such effects change the
total energy by less than 0.1~eV, and can therefore safely be disregarded.

We next set up the molecules in the SET electrostatic environment. The geometry is
illustrated in \ref{fig:benzene_set}. It consists of a metallic back-gate,
and above the gate there is 3.8~\AA\ of dielectric material with dielectric constant
$10 \epsilon_0$. The molecule is positioned 1.2~\AA\ above the
dielectric. To the left and right of the molecule are metallic
source-drain electrodes, and the distance between the molecule and the
electrodes is 2.8~\AA. We note that for a typical metal surface the
image plane is 2~\AA\ above the surface\cite{chulkov1999}, and to compare with atomic
adsorption geometries this length must be added to the above
distances.

\ref{tab:total_energy} lists the charging energies of the molecules in
the SET environment with zero potential at the gate electrode. We see
that for benzene the charging energy is strongly reduced, while the
effect is smaller for C$_{60}$. The reduction in charging energy arises from the screening
of the charged molecule by the surrounding dielectric and metal
electrodes\cite{Kaasbjerg2008}.

For benzene we may compare with the GW calculations by Neaton
{\textit et al.}\cite{Neaton2006}. They find the gas phase value for
$E_\mathrm{LUMO}-E_\mathrm{HOMO} = 10.51\ \mathrm{eV}$ and for benzene adsorbed on
graphene $E_\mathrm{LUMO}-E_\mathrm{HOMO} = 7.35\ \mathrm{eV}$, which is comparable
with our values of $E_\mathrm{I}-E_\mathrm{A}= 11.49\  \mathrm{eV}$ in the
gas phase and $E_\mathrm{I}-E_\mathrm{A}= 7.67\  \mathrm{eV}$ in the SET environment.

\begin{figure}[tbp]
\begin{center}
\includegraphics[width=\linewidth]{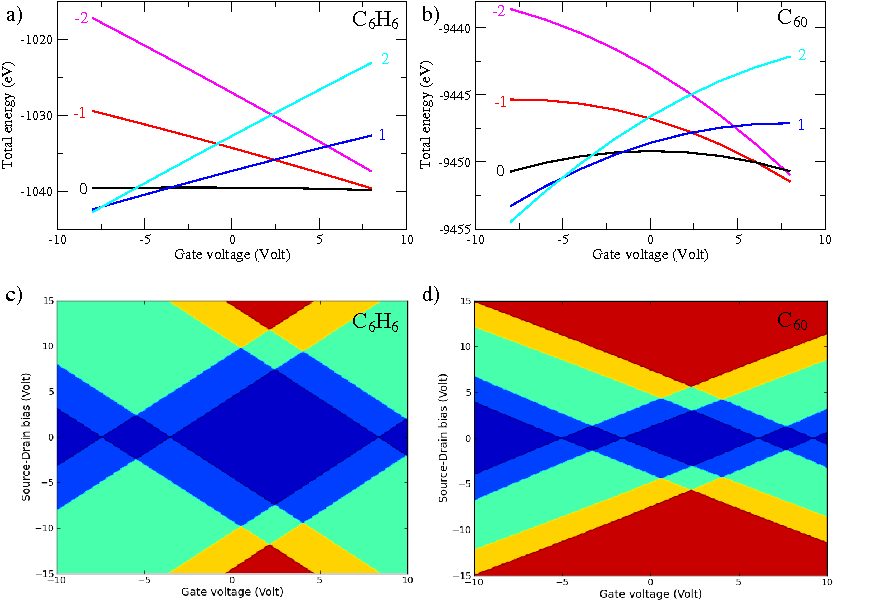}
\end{center}
\caption{(Color online) a) Total energy as function of the gate
  voltage for the benzene molecule, in the SET environment. Different
  curves are for charge states $-2$, $-1$, 0, 1, 2. c) Charge stability
  diagram; the color  shows the number of charge
  states within the bias window for a given gate voltage and
  source-drain bias. dark blue(0), blue(1), green(2), yellow(3), red(4).  b) and d) shows similar plots for the
  C$_{60}$ molecule, in the SET environment.}\label{fig:set-data}
\end{figure}

We next calculate the total energy of the different charge states
of the SET system as function of
the gate potential. The results are shown in \ref{fig:set-data}. The
total energy includes the reservoir energy $qW$, where $q$ is the
charge of the molecule and $W$ is the work function of the
electrode; we use the value $W=5.28\ \mathrm{eV}$ which models a gold
electrode\cite{AuW}.

\ref{fig:set-data} shows that the neutral molecule has the
lowest energy at zero gate potential. At negative gate potentials the
positive charge states are stabilized, while the negative charge states
are stabilized at positive gate potentials. This is in agreement with
that HOMO and LUMO levels follow $- e V_G$, thus, at positive bias the
LUMO level gets below the electrode Fermi level and attracts an
electron, and the molecule becomes negatively charged. At negative
gate potentials the HOMO level gets above the electrode Fermi level
and an electron is escaping from the molecule, which becomes
positively charged. 

To understand the dependence
between the total energy and the gate potential we fit a quadratic
function to the data
\begin{equation}
E = \alpha q V_G + \beta (e V_G)^2 .
\label{eq:gate_coupling}
\end{equation}
Note that we assume the linear term to be proportional to the charge $q$ on
the molecule, while the quadratic term arises from polarization of the
molecule and therefore is independent of $q$. By fitting the data
in \ref{fig:set-data} we find for benzene
$\alpha = 0.62\ \mathrm{eV}$, $ \beta=-0.003\  \mathrm{eV}^{-1} $, and for C$_{60}$ $\alpha = 0.38\  \mathrm{eV}$,
$\beta=-0.025\  \mathrm{eV}^{-1} $, where the variation with the charge
state is  $\sim0.01\  \mathrm{eV}$ for $\alpha$ and $\sim0.001\
\mathrm{eV}^{-1} $ for $\beta$.

Thus, benzene is stronger coupled with the gate than C$_{60}$, because
the benzene atoms on average are closer to the dielectric substrate.
Therefore, the benzene molecule shows an almost linear relationship between the total
energy and the gate potential, since all atoms are almost
identically shifted by the gate potential.

For C$_{60}$, on the other hand, the relationship between the total
energy and the gate potential is non-linear. For this molecule the atoms
closest to the dielectric region screen the gate potential for the rest of
the molecule, and thus the gate coupling becomes smaller. A difference in the charges on
different atoms in the molecule gives rise to a molecular dipole, and it is this
polarization energy that gives the second-order contribution to the
total energy.

From the total energies we can finally calculate the charge stability diagram
using \ref{eq:charge-stability}. The result is show in
\ref{fig:set-data}c,d. The different colors show the number of charge
states in the bias window. We see that the excitation energy for
C$_{60}$ is smaller than for benzene.  For both systems the excitation energy
of the second electron is much smaller than for the first electron.
We also note that the non-linear dependence of the total energy on the
gate potential for C$_{60}$ is not observable in the
charge stability diagram. This is because the charge stability diagram only depends on
the energy differences between the charge states, and the second-order term in
\ref{eq:gate_coupling} is independent on the charge state.

\section{Conclusions}

We have in this paper demonstrated the use of density-functional
theory for calculating the charging
energy of a molecule in a metallic environment that models the
geometry of a molecular single-electron transistor. We find that the
metallic environment reduces the charging energy of the molecules, in
agreement with GW\cite{Neaton2006} and H\"uckel\cite{Kaasbjerg2008} calculations.
We calculated the charging energy as function of the gate potential and from this
obtained the charge stability diagram. The simulations show how DFT
can be used to gain new insight into the properties of molecular
single-electron transistors operating in the coulumb blockade regime.

\acknowledgement
This work was supported by the Danish Council for Strategic Research 'NABIIT'
under Grant No.\ 2106-04-0017, ``Parallel Algorithms for Computational
Nano-Science'', and European Commission STREP project No.\ MODECOM
``NMP-CT-2006-016434'', EU. I would also like to thank Anders Blom for
important comments to the manuscript. 


\bibliography{paper}

\providecommand*{\mcitethebibliography}{\thebibliography}
\csname @ifundefined\endcsname{endmcitethebibliography}
{\let\endmcitethebibliography\endthebibliography}{}
\begin{mcitethebibliography}{25}
\providecommand*{\natexlab}[1]{#1}
\providecommand*{\mciteSetBstSublistMode}[1]{}
\providecommand*{\mciteSetBstMaxWidthForm}[2]{}
\providecommand*{\mciteBstWouldAddEndPuncttrue}
  {\def\EndOfBibitem{\unskip.}}
\providecommand*{\mciteBstWouldAddEndPunctfalse}
  {\let\EndOfBibitem\relax}
\providecommand*{\mciteSetBstMidEndSepPunct}[3]{}
\providecommand*{\mciteSetBstSublistLabelBeginEnd}[3]{}
\providecommand*{\EndOfBibitem}{}
\mciteSetBstSublistMode{f}
\mciteSetBstMaxWidthForm{subitem}{(\alph{mcitesubitemcount})}
\mciteSetBstSublistLabelBeginEnd{\mcitemaxwidthsubitemform\space}
{\relax}{\relax}

\bibitem[Lang(1995)]{Lang1995}
Lang,~N.~D. \emph{Phys. Rev. B} \textbf{1995}, \emph{52}, 5335\relax
\mciteBstWouldAddEndPuncttrue
\mciteSetBstMidEndSepPunct{\mcitedefaultmidpunct}
{\mcitedefaultendpunct}{\mcitedefaultseppunct}\relax
\EndOfBibitem
\bibitem[{Xue}(2002)]{Xue2002}
{Xue},~Y. \emph{Chemical Physics} \textbf{2002}, \emph{281}, 151--170\relax
\mciteBstWouldAddEndPuncttrue
\mciteSetBstMidEndSepPunct{\mcitedefaultmidpunct}
{\mcitedefaultendpunct}{\mcitedefaultseppunct}\relax
\EndOfBibitem
\bibitem[Brandbyge et~al.(2002)Brandbyge, Mozos, Ordej\'on, Taylor, and
  Stokbro]{Brandbyge-2002-PRB}
Brandbyge,~M.; Mozos,~J.-L.; Ordej\'on,~P.; Taylor,~J.; Stokbro,~K. \emph{Phys.
  Rev. B} \textbf{2002}, \emph{65}, 165401\relax
\mciteBstWouldAddEndPuncttrue
\mciteSetBstMidEndSepPunct{\mcitedefaultmidpunct}
{\mcitedefaultendpunct}{\mcitedefaultseppunct}\relax
\EndOfBibitem
\bibitem[Taylor et~al.(2001)Taylor, Guo, and Wang]{Taylor2001}
Taylor,~J.; Guo,~H.; Wang,~J. \emph{Phys. Rev. B} \textbf{2001}, \emph{63},
  245407\relax
\mciteBstWouldAddEndPuncttrue
\mciteSetBstMidEndSepPunct{\mcitedefaultmidpunct}
{\mcitedefaultendpunct}{\mcitedefaultseppunct}\relax
\EndOfBibitem
\bibitem[Magoga and Joachim(1997)]{MaJo97}
Magoga,~M.; Joachim,~C. \emph{Phys. Rev. B} \textbf{1997}, \emph{56},
  4722\relax
\mciteBstWouldAddEndPuncttrue
\mciteSetBstMidEndSepPunct{\mcitedefaultmidpunct}
{\mcitedefaultendpunct}{\mcitedefaultseppunct}\relax
\EndOfBibitem
\bibitem[Corbel et~al.(1999)Corbel, Cerd\'{a}, and Sautet]{CoCeSa99}
Corbel,~S.; Cerd\'{a},~J.; Sautet,~P. \emph{Phys. Rev. B} \textbf{1999},
  \emph{60}, 1989\relax
\mciteBstWouldAddEndPuncttrue
\mciteSetBstMidEndSepPunct{\mcitedefaultmidpunct}
{\mcitedefaultendpunct}{\mcitedefaultseppunct}\relax
\EndOfBibitem
\bibitem[Cerd\'a and Soria(2000)]{Cerda2000}
Cerd\'a,~J.; Soria,~F. \emph{Phys. Rev. B} \textbf{2000}, \emph{61},
  7965--7971\relax
\mciteBstWouldAddEndPuncttrue
\mciteSetBstMidEndSepPunct{\mcitedefaultmidpunct}
{\mcitedefaultendpunct}{\mcitedefaultseppunct}\relax
\EndOfBibitem
\bibitem[Emberly and Kirczenow(2001)]{EmKi01}
Emberly,~E.~G.; Kirczenow,~G. \emph{Phys. Rev. B} \textbf{2001}, \emph{62},
  10451\relax
\mciteBstWouldAddEndPuncttrue
\mciteSetBstMidEndSepPunct{\mcitedefaultmidpunct}
{\mcitedefaultendpunct}{\mcitedefaultseppunct}\relax
\EndOfBibitem
\bibitem[Zahid et~al.(2005)Zahid, Paulsson, Polizzi, Ghosh, Siddiqui, and
  Datta]{Zahid2005}
Zahid,~F.; Paulsson,~M.; Polizzi,~E.; Ghosh,~A.~W.; Siddiqui,~L.; Datta,~S.
  \emph{J. of Chem. Phys.} \textbf{2005}, \emph{123}, 064707\relax
\mciteBstWouldAddEndPuncttrue
\mciteSetBstMidEndSepPunct{\mcitedefaultmidpunct}
{\mcitedefaultendpunct}{\mcitedefaultseppunct}\relax
\EndOfBibitem
\bibitem[Kienle et~al.(2006)Kienle, Cerd\'{a}, and Ghosh]{KienleI2006}
Kienle,~D.; Cerd\'{a},~J.~I.; Ghosh,~A.~W. \emph{J. Appl. Phys.} \textbf{2006},
  \emph{100}, 043714\relax
\mciteBstWouldAddEndPuncttrue
\mciteSetBstMidEndSepPunct{\mcitedefaultmidpunct}
{\mcitedefaultendpunct}{\mcitedefaultseppunct}\relax
\EndOfBibitem
\bibitem[Kienle et~al.(2006)Kienle, Bevan, Liang, Siddiqui, Cerd\'{a}, and
  Ghosh]{KienleII2006}
Kienle,~D.; Bevan,~K.~H.; Liang,~G.-C.; Siddiqui,~L.; Cerd\'{a},~J.~I.;
  Ghosh,~A.~W. \emph{J. Appl. Phys.} \textbf{2006}, \emph{100}, 043715\relax
\mciteBstWouldAddEndPuncttrue
\mciteSetBstMidEndSepPunct{\mcitedefaultmidpunct}
{\mcitedefaultendpunct}{\mcitedefaultseppunct}\relax
\EndOfBibitem
\bibitem[Stokbro et~al.()Stokbro, Petersen, Smidstrup, Blom, Ipsen, and
  Kaasbjerg]{stokbro2010}
Stokbro,~K.; Petersen,~D.~E.; Smidstrup,~S.; Blom,~A.; Ipsen,~M.; Kaasbjerg,~K.
  Submitted, arXiv:1004.2812v1 (\url{http://arxiv.org/abs/1004.2812})\relax
\mciteBstWouldAddEndPuncttrue
\mciteSetBstMidEndSepPunct{\mcitedefaultmidpunct}
{\mcitedefaultendpunct}{\mcitedefaultseppunct}\relax
\EndOfBibitem
\bibitem[Kubatkin et~al.(2003)Kubatkin, Danilov, Hjort, Cornil, Bredas,
  Stuhr-Hansen, Hedeg{\aa}rd, and Bj{\o}rnholm]{bjornholm}
Kubatkin,~S.; Danilov,~A.; Hjort,~M.; Cornil,~J.; Bredas,~J.-L.;
  Stuhr-Hansen,~N.; Hedeg{\aa}rd,~P.; Bj{\o}rnholm,~T. \emph{Nature}
  \textbf{2003}, \emph{425}, 698\relax
\mciteBstWouldAddEndPuncttrue
\mciteSetBstMidEndSepPunct{\mcitedefaultmidpunct}
{\mcitedefaultendpunct}{\mcitedefaultseppunct}\relax
\EndOfBibitem
\bibitem[ATK()]{ATK2010.02}
 Atomistix ToolKit version 2010.02, QuantumWise A/S
  (\texttt{http://quantumwise.com/})\relax
\mciteBstWouldAddEndPuncttrue
\mciteSetBstMidEndSepPunct{\mcitedefaultmidpunct}
{\mcitedefaultendpunct}{\mcitedefaultseppunct}\relax
\EndOfBibitem
\bibitem[Soler et~al.(2002)Soler, Artacho, Gale, Garc\'{i}a, Junquera,
  Ordej\'{o}n, and S\'{a}nchez-Portal]{SIESTA2002}
Soler,~J.~M.; Artacho,~E.; Gale,~J.~D.; Garc\'{i}a,~A.; Junquera,~J.;
  Ordej\'{o}n,~P.; S\'{a}nchez-Portal,~D. \emph{Journal of Physics: Condensed
  Matter} \textbf{2002}, \emph{14}, 2745--2779\relax
\mciteBstWouldAddEndPuncttrue
\mciteSetBstMidEndSepPunct{\mcitedefaultmidpunct}
{\mcitedefaultendpunct}{\mcitedefaultseppunct}\relax
\EndOfBibitem
\bibitem[Neugebauer and Scheffler(1992)]{NeSc92}
Neugebauer,~J.; Scheffler,~M. \emph{Phys. Rev. B} \textbf{1992}, \emph{46},
  16067\relax
\mciteBstWouldAddEndPuncttrue
\mciteSetBstMidEndSepPunct{\mcitedefaultmidpunct}
{\mcitedefaultendpunct}{\mcitedefaultseppunct}\relax
\EndOfBibitem
\bibitem[Perdew and Zunger(1981)]{Perdew1981}
Perdew,~J.~P.; Zunger,~A. \emph{Phys. Rev. B} \textbf{1981}, \emph{23},
  5048--5079\relax
\mciteBstWouldAddEndPuncttrue
\mciteSetBstMidEndSepPunct{\mcitedefaultmidpunct}
{\mcitedefaultendpunct}{\mcitedefaultseppunct}\relax
\EndOfBibitem
\bibitem[Lias et~al.(1988)Lias, Bartmess, Liebman, Holmes, Levin, and
  Mallard]{benzene_ion}
Lias,~S.~G.; Bartmess,~J.~E.; Liebman,~J.~E.; Holmes,~J.~L.; Levin,~R.~D.;
  Mallard,~W.~G. \emph{J. Phys. Chem. Ref. Data} \textbf{1988}, \emph{17 (suppl
  1)}, year\relax
\mciteBstWouldAddEndPuncttrue
\mciteSetBstMidEndSepPunct{\mcitedefaultmidpunct}
{\mcitedefaultendpunct}{\mcitedefaultseppunct}\relax
\EndOfBibitem
\bibitem[Pogulay et~al.(2004)Pogulay, Abzalimov, Nasibullaev, Lobach, Drewello,
  and Vasil\'{e}v]{c60_ion}
Pogulay,~A.~V.; Abzalimov,~R.~R.; Nasibullaev,~S.~K.; Lobach,~A.~S.;
  Drewello,~T.; Vasil\'{e}v,~Y.~V. \emph{Int. J. of Mass Spec.} \textbf{2004},
  \emph{233}, 165\relax
\mciteBstWouldAddEndPuncttrue
\mciteSetBstMidEndSepPunct{\mcitedefaultmidpunct}
{\mcitedefaultendpunct}{\mcitedefaultseppunct}\relax
\EndOfBibitem
\bibitem[Tosatti and Manini(1994)]{c60_a}
Tosatti,~E.; Manini,~N. \emph{Chem. Phys. Lett.} \textbf{1994}, \emph{223},
  61\relax
\mciteBstWouldAddEndPuncttrue
\mciteSetBstMidEndSepPunct{\mcitedefaultmidpunct}
{\mcitedefaultendpunct}{\mcitedefaultseppunct}\relax
\EndOfBibitem
\bibitem[Chulkov et~al.(1999)Chulkov, Silkin, and Echenique]{chulkov1999}
Chulkov,~E.; Silkin,~V.; Echenique,~P. \emph{Surface Science} \textbf{1999},
  \emph{437}, 330\relax
\mciteBstWouldAddEndPuncttrue
\mciteSetBstMidEndSepPunct{\mcitedefaultmidpunct}
{\mcitedefaultendpunct}{\mcitedefaultseppunct}\relax
\EndOfBibitem
\bibitem[Kaasbjerg and Flensberg(2008)]{Kaasbjerg2008}
Kaasbjerg,~K.; Flensberg,~K. \emph{Nano Lett.} \textbf{2008}, \emph{8},
  3809\relax
\mciteBstWouldAddEndPuncttrue
\mciteSetBstMidEndSepPunct{\mcitedefaultmidpunct}
{\mcitedefaultendpunct}{\mcitedefaultseppunct}\relax
\EndOfBibitem
\bibitem[Neaton et~al.(2006)Neaton, Hybertsen, and Louie]{Neaton2006}
Neaton,~J.~B.; Hybertsen,~M.~S.; Louie,~S.~G. \emph{Phys. Rev. Lett.}
  \textbf{2006}, \emph{97}, 216405\relax
\mciteBstWouldAddEndPuncttrue
\mciteSetBstMidEndSepPunct{\mcitedefaultmidpunct}
{\mcitedefaultendpunct}{\mcitedefaultseppunct}\relax
\EndOfBibitem
\bibitem[Rivi\'{e}re(1966)]{AuW}
Rivi\'{e}re,~J.~C. \emph{Appl. Phys. Lett.} \textbf{1966}, \emph{8}, 172\relax
\mciteBstWouldAddEndPuncttrue
\mciteSetBstMidEndSepPunct{\mcitedefaultmidpunct}
{\mcitedefaultendpunct}{\mcitedefaultseppunct}\relax
\EndOfBibitem
\end{mcitethebibliography}

\end{document}